# Highlights of AAC 2000 Workshop

J.B. Rosenzweig, Department of Physics and Astronomy,
University of California, Los Angeles, 405 Hilgard Ave., Los Angeles, CA 90095

*Abstract*

The Advanced Accelerator Concepts 2000 (AAC2K) Workshop was held in Santa Fe in June, 2000, and included a wide array of conceptual and theoretical advances at the frontier of accelerator physics. This paper reviews the highlights of the workshop, with subjects ranging from acceleration using lasers, plasmas and microstructures, to the beam physics of muon colliders. Particular emphasis is given to the topics which are relevant to research at existing linear accelerator facilities, and the effect of this research on the capabilities of such facilities.

## 1 INTRODUCTION

The field of advanced accelerators, which is in search of new and revolutionary technology to allow progress in high energy physics experimentation, holds approximately biannual meetings. The latest in this series, Advanced Accelerator Concepts 2000, was held in June, 2000 in Santa Fe, New Mexico, and was hosted by Los Alamos National Laboratory, which also hosted the original workshop in the series[1].

The advanced accelerator field initially emerged in the early 1980's as a response to the need to scale linac technology to TeV-class colliders. It was realized at that time, and it is still true today, that some dramatic change in the physical and technological paradigms of charged acceleration are necessary in order to build a compact and affordable version of such a collider. The search for such alternate technologies has now matured into a vigorous sub-field of accelerator physics. The field of advanced accelerator concepts (AAC) has renewed sense of purpose in light of both the failure of the SSC, and in the considerable experimental progress made in the field.

In this paper, we review the conceptual structure of the AAC field, and mark its progress as of AAC2K. It is important to keep in mind that, even with dramatic proof-of-principle experimental results, that AAC has yet to produce a working accelerator in the sense that the linac community would recognize one. Nevertheless, the AAC field has already made a strong impact on the linac field, as AAC efforts have pushed the state of the art in linac-based experiments. AAC work has thus stimulated great progress in electron beam sources (notably the rf photo-injector), and in ultra-low emittance, sub-picosecond beam measurements. These techniques have already found their way into the more conventional fields of linear colliders and free-electron lasers. It can be expected that AAC work will next introduce non-conventional technologies into the rf linac and related high energy physics fields, in the form of novel radiation and particle sources, and lenses.

## 2 AAC: PHYSICAL PRINCIPLES

In order to review the progress made in AAC that was evident in the workshop, we first introduce the major categories of advanced accelerator schemes.

- Wake-field accelerators (WFA): This type of accelerator uses what might be best termed a novel approach to creation of high frequency rf power, using a tightly bunched beam pulse or train of pulses which traverses a high impedance environment. The accelerating beam may traverse the same environment (collinear WFA) or a nearby structure (a two-beam accelerator, or TBA). The WFA addressed the problems of power creation and distribution at high frequency and high fields. The highest impedance environment one may use is a plasma (PWFA[2]) of high density $n_0$, in which the power radiated by the beam in its wake (generalized Cerenkov radiation) is in the form of electrostatic or electromagnetic plasma waves. On the more conventional side, dielectric and metallic structures have been studied for WFA use; the CLIC[3] linear collider is essentially a TBA based on metallic structures.
- Direct laser acceleration: Since lasers are known to be a cheap and efficient source of electromagnetic radiation at extremely high field and power levels, they are very attractive for AAC applications. For relativistic beam particles, however, acceleration is only made possible by bending particle trajectories (inverse free-electron laser, or IFEL[4]), or by introducing non-vacuum boundary or impedance conditions. These conditions can be similar to rf linac structures, or deformed into a planar geometry[5,6], or use the inverse Cerenkov effect, or ICA[7]. The obstacles to realization of such schemes center on the problem of scaling the accelerating wave down in size by four orders of magnitude(!) from present rf linac technology. On the other hand, laser-based techniques have the tide of history on their side; all technologies have been pushed towards miniaturization in recent years.
- Plasma accelerators: This classification clearly overlaps with the previous two, but is often discussed separately in the context of intense laser-plasma interaction. Plasma waves can be excited by lasers, just as with electron beams in the case of the PWFA, when the laser pulse is very short compared to the plasma wavelength $\lambda_p = \sqrt{\pi/n_0 r_e}$ (laser wake-field accelerator, or LWFA[8]) and modulated at the plasma wavelength (plasma beatwave accelerator, or PBWA[9]). In plasmas, since the medium is already ionized, structure breakdown does not limit the field amplitude, and acceleration rates of several GeV/m have been reported[8,9]. Plasma accelerators have sev-eral interesting byproducts, notably plasma lenses[10], which may have application in linear collider final foci, and ultra-high brightness particle sources[11].

## 3 PROGRESS REPORTED AT AAC2K

### 3.1 Wake-field Acceleration

Progress in WFA was reported in a number of significant areas at AAC2K. Investigators at the facility most dedicated to WFA research, the Argonne Wake-field Accelerator (AWA), showed experimental results in the areas of multiple-pulse excitation[12] which verified the linear theory of Cerenkov wake-fields. In addition, the AWA performed initial acceleration experiments using the so-called step-up transformer, a form TBA in which the rf power excited by the drive beam wake in one dielectric tube is transferred to another tube of higher impedance, in which a test beam is accelerated[13]. The investigation of dielectric wakes (coherent Cerenkov emission) has been extended at the AWA to an experiment designed to observe a signal similar to that expected by an EM shower induced by the interaction of a high energy neutrino with lunar matter[14], as in a newly proposed scheme of ultra-high energy neutrino detection.

The E-157 collaboration has reported acceleration using the PWFA mechanism in the so-called "blow-out"[2] regime at the SLAC FFTB. The 3.3 nC, 2 psec rms, 30 GeV beam is injected into a 1.4 m long, very uniform plasma of density near $10^{14}$ cm$^{-3}$. Much of the beam is decelerated, with the tail being accelerated by the wake-field. This beam was observed by a time-resolved imaging and energy measurement system. The analysis of the data taken in this experiment has proven difficult due to the presence of strong transverse kicks which mimic energy changes in the spectrometer. The data is, despite these problems, consistent with the computational models of the PWFA in this nonlinear regime, with wakes in the range of several hundred MeV/m measured[15].

As the acceleration gradient in wake-field accelerators has long been recognized to scale strongly with the bunch length (with $\sigma_z^{-2}$), several AAC photoinjector labs have developed bunch compressors. Pulse compression with such high currents and low emittances has inherent physics interest, due to such poorly understood processes as coherent synchrotron radiation and related emittance growth. Also measurement of such pulses in the sub-psec regime presents serious challenges, giving rise to methods beyond the limits of streak camera resolution[16] which rely on coherent transition radiation. The performance of compressor systems at the FNAL A0 photoinjector[16], the UCLA Neptune lab[17], and the Univ. of Tokyo[18], were all reported at AAC2K. The 5 nC, 2 psec bunches produced at A0 were used by a UCLA team to drive wake-fields in the blow-out regime which also reached several hundred MeV/m, nearly stopping the 17 MeV beam in 8 cm of a $7\times10^{13}$ cm$^{-3}$ plasma[19].

The availability of 50 GeV electron beams which can be compressed to less than 30 μm rms pulse length at the end of the SLC arcs at SLAC gave rise to the suggestion by Katsouleas that a dense plasma of 5 m length placed after the arcs could double the energy of the SLC collisions[20]. This suggestion, termed the wake-field "after-burner", will undoubtedly spur further investigation.

### 3.2 Laser Acceleration

Experiments on direct laser acceleration which use structures, despite having been generally proposed for over a decade, have just begun. The daunting challenges (aperture, timing) of scaling the acceleration wavelength into the infrared have proven to be considerable. Progress in a 1 μm(!) wave-length planar-geometry experiment termed LEAP[21] at Stanford was reported at AAC2K, yet without definitive proof of acceleration .

Acceleration, however, has been reported for years in non-structure-based (IFEL, ICA) experiments at the BNL ATF using 10 μm light. The next generation of laser acceleration experiment, in the form of a scheme termed STELLA, has just begun to produce impressive results as reported at AAC2K. STELLA (STaged ELectron Laser Acceleration[22]) is a two-stage system based on a 10 μm IFEL. The first stage microbunches the several psec (mm) ATF beam into the 10 μm IFEL period. This beam is then injected at the correct phase for acceleration into the second stage , where the captured beam is accelerated with small phase and energy spread. Some results of this remarkable experiment are shown in Fig. 1, which displays the momentum spectrometer images of a beam before interaction, after the bunching of the beam in the first stage, and after the capture and acceleration of the beam in the second IFEL stage. This robust result thus shows the first precise manipulation of accelerating electron beams in waves with only 31 fsec period.

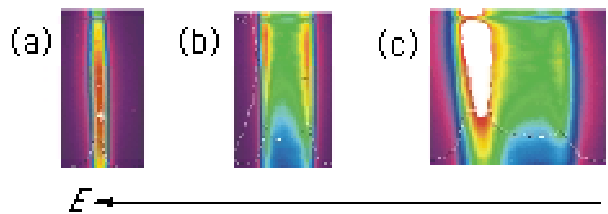

Figure 1. False-color spectrometer images from STELLA experiment at BNL ATF (a) without IFEL, (b) after initial stage IFEL bunching, and (c) after correctly phased second IFEL stage (courtesy Ilan Ben-zvi).

### 3.3 Plasma Accelerators, Lenses and Sources

In the past few AAC workshops, great leaps forward into multi-GeV/m acceleration was reported in laser-driven plasma accelerators. No such discoveries were reported at AAC2K, which found other plasma-based experiments, including PWFA, in the spotlight.

In the context of PWFA, an important milestone in beam-plasma interaction was reported by the E-150

collaboration at SLAC. In this experiment, positron beams at the FFTB were injected into short plasmas, with the measured focusing of the beam from 8 μm to 4 μm rms width[10,23]. This promising result has resounding implications for use of plasmas for focusing and accelerating positrons in future colliders.

The highest level of attention at AAC2K in the area of laser-plasma acceleration was given to the production of high-brightness electron beams by the breaking of large amplitude, laser-driven plasma waves. Pioneering work on this effect performed at the Univ. of Michigan (UM) was reviewed by Umstadter[24]. Further work using a more intense laser at LBNL was also presented, in which the extracted high brightness fsec electron pulses were measured to have energies in excess of 25 MeV[25]. Further control over the injection dynamics of these electron micropulses is expected when the wave-breaking is dictated by the collision of the a second laser pulse[25], or caused by a sharp density transition, as proposed in the context of the PWFA by Suk, et al.[26].

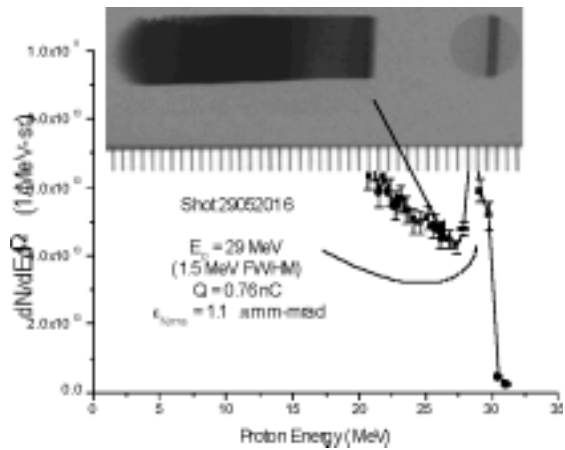

Figure 2. Radiographic film spectrum of protons emitted by PW laser interaction with solid target, showing peak near maximum energy (courtesy T. Cowan).

*3.4 Heavy Particle Acceleration*

While the AAC Workshop traditionally is focused on acceleration and manipulation of electrons and positrons, at AAC2K there was much discussion of advances in heavy particle accelerator physics.

The recent emphasis of laser-plasma based sources of electrons has been recently paralleled by the unexpected discovery of ion beam generation from multi-TW to PW, sub-psec laser-solid interactions. At AAC2K, results from the UM group (in the ten TW regime), as well as from LLNL (PW class laser) were reported. In the lower power experiments, it has been shown that the energetic (several MeV) ions are generated on the upstream sided of the thin-foil targets. In the LLNL experiments[27], on the other hand, the accelerated ions (>30 MeV) are found to be derived from the back side of the target, ejected along the surface normal. In both experiments, the emitted ions showed small transverse phase space extent. Even more remarkably, in the LLNL case the energy spectrum showed a notable peak near the maximum observed energy — the emitted ions formed a true narrow band (and initially sub-picosecond) beam. While such observation of such unprecedentedly high brightness ion beam sources capture the imagination, it is not yet clear how one captures such a beam into a more conventional linac to use it in applications.

The recent emergence of muon colliders, and the related concept of neutrino factories, was recognized in the working groups at AAC2K. There are many challenging research problems which need to be investigated before such projects are undertaken, such as ionization cooling, and manipulation of beam phase spaces under conditions where the angular momentum of the beam is nonzero[28]. It was apparent from the presentations at AAC2K he muon/neutrino factory field is still in its youth, and that with the onset of active experiments to prove the relevant principles, that this proposed high-energy physics instrument will grow in importance in the near future.

## 4 REFERENCES

1. *Laser Acceleration of Particles,* AIP Conf. Proc. **9 1** (AIP, New York, 1982)
2. J.B.Rosenzweig, *et al.*, *Phys.Rev.A* **44**, R6189 (1991)
3. "The CLIC Study of a Multi-TeV e± Linear Collider", Jean-Pierre Delahaye, *et al. Proc. 1999 Part. Accel. Conf.* 250 (IEEE, New York,1999)
4. R. Palmer, *J. Appl. Phys.,* **43** 3014 (1972).
5. Y.Huang, R.L.Byer, *Appl.Phys. Lett.* **69** 2175 (1996)
6. J.B. Rosenzweig, A. Murokh, and C. Pellegrini, *Phys. Rev. Lett.* **74**, 2467 (1995).
7. W.D. Kimura, *et al., Phys. Rev. Lett*. **74**, 546 (1995*)*
8. K. Nakajima, *et al., Phys. Rev. Lett.* **74**, 4428 (1995)
9. C.E. Clayton, *et al., Phys. Rev. Lett.* **70**, 37 (1993)
10. P. Chen, *et al. Phys. Rev. D* **40**, 923-926 (1989)
11. D.Umstadter *et al., Phys. Rev. Lett.* 76, 2073 (1996)
12. J.G. Power, *et al., Phys. Rev. E,* **60**, 6061 (1999).
13. P. Zou, *et al., Rev. Sci. Inst.* **71,** 2301 (2000).
14. P. Gorham, *et al.,* LANL e-print hep- ex/ 0004007.
15. P. Muggli, these proceedings.
16. M. Fitch, *et al. . Proc. 1999 Part. Accel. Conf.* 2181 (IEEE, New York,1999).
17. S. Anderson, *et al.,* to appear in *Proc. AAC2K* (AIP)
18. M. Uesaka, *et al.,* to appear in *Proc. AAC2K* (AIP)
19. N. Barov, *et al.,* to appear in *Proc. AAC2K* (AIP)
20. T. Katsouleas, to appear in *Proc. AAC2K* (AIP)
21. E. Colby, *et al.,* to appear in *Proc. AAC2K* (AIP)
22. W.D Kimura, *et al.,* to appear in *Proc. AAC2K* (AIP)
23. J. Ng, *et al.,* to appear in *Proc. AAC2K* (AIP)
24. Xiaofang Wang, *et al. PRL,* **84**,5324 (2000)
25. W. Leemans, these proceedings.
26. H. Suk, *et al.,* to appear in *Proc. AAC2K* (AIP)
27. T. Cowan, to appear in *Physics of High Brightness Beams* (World Sci., Singapore, 2000).
28. G.Penn, J.S.Wurtele, *Phys.Rev.Lett*. **85**, 764 (2000)